\documentclass[aps,prl,twocolumn,superscriptaddress,showpacs,10pt]{revtex4-1}

\usepackage{graphicx}

\usepackage{amsmath}

\usepackage{color}

\usepackage{multirow,bigstrut}

\usepackage{time,mathptmx}

\definecolor{darkblue}{rgb}{0.22,0.33,0.64}

\usepackage[
pdfstartview=XYZ,
bookmarks=false,
colorlinks=true,
linkcolor=darkblue,
urlcolor=darkblue,
citecolor=darkblue
]{hyperref}

\makeatletter

\newcommand{\Rmnum}[1]{\expandafter\@slowromancap\romannumeral #1@}
\makeatother

\begin{document}

\title{Precision Mapping of Laser-Driven Magnetic Fields and their \\Evolution in High-Energy-Density Plasmas}

\author{L.~Gao}
\altaffiliation{Current address: Department of Astrophysical Sciences, Princeton University, Princeton, NJ, 08544, USA}
\affiliation{Laboratory for Laser Energetics, University of Rochester, Rochester, NY, 14623, USA} 
\affiliation{Department of Mechanical Engineering, University of Rochester, Rochester, NY, 14623, USA}

\author{P.~M.~Nilson}
\affiliation{Laboratory for Laser Energetics, University of Rochester, Rochester, NY, 14623, USA} 
\affiliation{Fusion Science Center for Extreme States of Matter, University of Rochester, Rochester, NY, 14623, USA}

\author{I.~V.~Igumenschev}
\affiliation{Laboratory for Laser Energetics, University of Rochester, Rochester, NY, 14623, USA} 

\author{M.~G.~Haines}
\altaffiliation{Deceased.}
\affiliation{Department of Physics, Imperial College, London SW7 2AZ United Kingdom}

\author{D.~H.~Froula}
\affiliation{Laboratory for Laser Energetics, University of Rochester, Rochester, NY, 14623, USA}

\author{R.~Betti}
\affiliation{Laboratory for Laser Energetics, University of Rochester, Rochester, NY, 14623, USA} 
\affiliation{Department of Mechanical Engineering, University of Rochester, Rochester, NY, 14623, USA}
\affiliation{Fusion Science Center for Extreme States of Matter, University of Rochester, Rochester, NY, 14623, USA}
\affiliation{Department of Physics and Astronomy, University of Rochester, Rochester, NY, 14623, USA}

\author{D.~D.~Meyerhofer}
\affiliation{Laboratory for Laser Energetics, University of Rochester, Rochester, NY, 14623, USA} 
\affiliation{Department of Mechanical Engineering, University of Rochester, Rochester, NY, 14623, USA}
\affiliation{Fusion Science Center for Extreme States of Matter, University of Rochester, Rochester, NY, 14623, USA}
\affiliation{Department of Physics and Astronomy, University of Rochester, Rochester, NY, 14623, USA}



\begin{abstract}

\noindent The magnetic fields generated at the surface of a laser-irradiated planar solid target were mapped using ultrafast proton radiography. Thick (50-$\mu$m) plastic foils were irradiated with 4-kJ, 2.5-ns laser pulses focused to an intensity of 4 $\times$ 10$^{14}$ W$/$cm$^{2}$. The data show magnetic fields concentrated at the edge of the laser-focal region, well within the expanding coronal plasma. The magnetic field spatial distribution was tracked and shows good agreement with 2-D resistive magnetohydrodynamic simulations using the code DRACO when the Biermann battery source, fluid and Nernst advection, resistive magnetic diffusion, and Righi-Leduc heat flow are included.

\end{abstract}




\maketitle

Large magnetic fields can be generated in electrically conducting fluids by a variety of mechanisms, including spatially non-uniform energy absorption \cite{Stamper1971,StamperRipin1975,Raven1978,Willingale2010} and hydrodynamic \cite{Mima1978,Manuel2012,Srinivasan_PRL_2012, Gao2012,Gao2013}, thermal \cite{Haines1981}, and thermomagnetic instabilities \cite{tidman:1207}. Much progress has been made in recent years on understanding the generation and transport of these fields in high-energy-density plasmas \cite{DrakeHEDP}, motivated by problems in laboratory astrophysics \cite{Remington28051999}, magnetic reconnection \cite{Nilson_MR,li2007observation,FoxMR,ZhongMR,fiksel2014magnetic}, hydrodynamic-instability growth \cite{Evans1986}, shock-wave dynamics \cite{Gregori_Nature,Fox2013,Meinecke}, and inertial confinement fusion \cite{Rygg2008}. In these conditions, much attention has been given to understanding the Biermann battery mechanism \cite{Biermann1951} and the magnetic fields that can be generated at the surface of a laser-irradiated solid target \cite{Haines1986}. 

The Biermann battery mechanism derives from the electron pressure gradient term in Ohm's law \cite{Biermann1951}. For a single laser beam focused on a solid target, non-collinear electron temperature and density gradients drive the magnetic-field growth \cite{Stamper1991}. The generated magnetic fields are azimuthal in their orientation \cite{cecchetti2009} and grow at the expense of the electron energy \cite{Haines1986}. Space- and time-resolved measurements showing where in the plasma these magnetic fields exist are important because they provide a stringent test for magnetohydrodynamic (MHD) model predictions \cite{Li2006,Li2007,Li2009,Petrasso2009,Nicolai2000}. The challenge is to accurately map the magnetic-field spatial distribution relative to the expanding coronal plasma. 

\begin{figure}[t]
\includegraphics{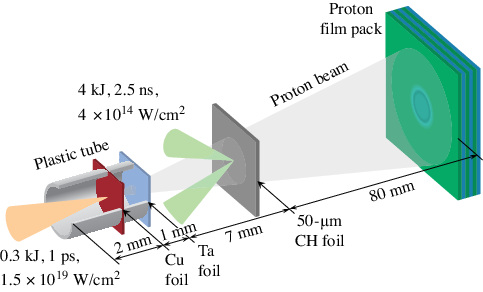}
\caption{\label{fig1} (color online). Experimental setup.}
\end{figure}

In this Letter, ultrafast proton radiography measurements of magnetic-field generation and transport at the surface of a laser-irradiated solid target are reported. The proton radiography data show magnetic fields not where they were previously thought to be. Previous work showed magnetic fields at the surface of a solid target concentrated on a hemispherical shell surrounding the laser-ablated plasma (or, `bubble'), with the maximum field amplitude near the bubble edge, falling to zero at its center \cite{Li2006,Li2007,Li2009,Petrasso2009}. The data reported here show magnetic fields concentrated at the edge of the laser-focal region, well within the expanding coronal plasma. These magnetic fields expand across the target surface at a speed smaller than the plasma sound speed in good agreement with 2-D resistive MHD simulations using the code DRACO \cite{radha2005}. The Biermann battery source \cite{Biermann1951}, fluid and Nernst advection \cite{nishiguchi1984convective, nishiguchi1985}, resistive magnetic diffusion \cite{Haines1986}, and Righi-Leduc heat flow \cite{Kho1985} had to be included in the calculations to reproduce the experimental measurements.

The experiments were carried out on the OMEGA EP Laser System \cite{Waxer2005} at the University of Rochester's Laboratory for Laser Energetics. Figure~\ref{fig1} shows a diagram of the experimental setup. OMEGA EP delivered two long pulse beams at a wavelength of 351-nm, each with $\sim$2-kJ of energy in a 2.5-ns square temporal profile. The long pulse beams were focused to 820-$\mu$m-diam focal spots at 23$^{\circ}$ angle of incidence to the target. The main targets were 50-$\mu$m-thick plastic foils, 5 $\times$ 5 mm$^{2}$ in area. The overlapped laser intensity was 4 $\times$ 10$^{14}$ W$/$cm$^{2}$. Each laser beam included distributed phase plates \cite{Lin1995}.

The proton radiography setup was the same as that described in Ref. \cite{Gao2013}. The radiography beam was an ultrafast laser-driven proton source generated from a 20-$\mu$m-thick Cu foil irradiated at normal incidence with a 0.3-kJ, 1-ps pulse. The high-intensity pulse had a wavelength of 1.053-$\mu$m and was focused with a 1-m focal length, f/2 off-axis parabolic mirror to an intensity of 1.5 $\times$ 10$^{19}$ W$/$cm$^{2}$. The high-energy proton beam was generated by target normal sheath acceleration \cite{wilks2001}. The Cu foil was mounted inside a plastic tube that was capped with a 5-$\mu$m-thick Ta foil, protecting the high-intensity interaction from the coronal plasma and x-ray pre-heat generated by the main target interaction \cite{zylstra:013511}.

\begin{figure}[t]
\includegraphics{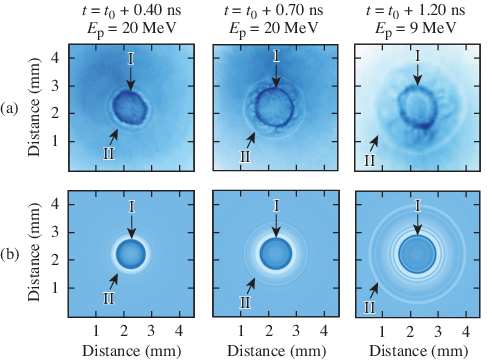}
\caption{\label{fig2}(color online). (a) Proton radiographs of 50-$\mu$m thick CH foils taken at $t=t_{0}+0.40$ ns, $t=t_{0}+0.70$ ns, and $t=t_{0}+1.20$ ns. The proton energy $E_{\rm p}$ used to generate each image is shown. (b) Synthetic proton radiographs calculated from the DRACO simulations (see text for details).}
\end{figure}

The main target was probed with up to several tens of MeV protons in a face-on geometry (see Fig.~\ref{fig1}). The protons were detected with a filtered stack of radiochromic film, providing two-dimensional images of the interaction \cite{Borghesi_ppcf_ProtonImaging}. Each film layer diagnosed the long-pulse interaction at different times based on the proton time-of-flight to the main target, the timing difference between the long- and short-pulse beams, and the energy dependent Bragg peak for proton energy deposition inside the detector. The system magnification was $M=(L + l)/l$, where $l$ was the distance from the proton-source foil to the long-pulse interaction and $L$ was the distance from the long-pulse interaction to the radiochromic film detector. For these experiments, $l$ was 8 mm and $L$ was between 80 and 96 mm, giving a system magnification $M$ $\sim$ 11 to 13. High spatial and temporal resolution was obtained in the 5- to 10-$\mu$m and few-picosecond range.

Figure~\ref{fig2}(a) shows proton radiographs of the spontaneous field structures that grew while the main laser was on. Data are shown from different shots at times $t=t_{0}+0.40$ ns, $t=t_{0}+0.70$ ns, and $t=t_{0}+1.20$ ns, where $t_{0}$ is the arrival time of the long-pulse beams at the main target surface. The effective integration time for each radiograph is a few picoseconds. The radiographic view of the target corresponded to 4.5 $\times$ 4.5 mm$^{2}$. The data show two circular structures growing in time. The inner circular structure (region I) was dark, indicating a higher detected proton flux compared to the outer circular structure (region II) that appeared as a lighter-colored ring. The inner dark ring was almost static, while the diameter of the outer light ring expanded at $\sim$1 $\times$ 10$^{8}$ cm$/$s. 





To understand the underlying source and transport of the fields that formed the light and dark circular structures, the data were compared to numerical model predictions from the 2D resistive MHD code DRACO \cite{radha2005}. In this model, the magnetic field evolves according to

\begin{equation}\label{eq:draco}
\frac{\partial \mathbf{B}} {\partial t}=\nabla\!\times\!\left\{\!\mathbf{u}\!\times\!\mathbf{B}+\frac{c}{e}\!\left[\frac{\nabla\!p_{e}}{n_{e}}-\frac{(\nabla\!\times\!\mathbf{B})\!\times\!\mathbf{B}}{4\pi n_{e}}-\frac{\mathbf{R}}{n_{e}}\right]\!\!\right\}\!,
\end{equation}

\mbox{}

\noindent where $\mathbf{u}$ is the flow velocity, $c$ is the speed of light,  $e$ is the fundamental unit of charge, $p_{e}$ is the electron pressure, $n_{e}$ is the electron density, and $\mathbf{R}=\mathbf{R_{T}}+\mathbf{R_{u}}$ includes the thermal and frictional forces \cite{Braginskii1965}. In this system, an azimuthal (Biermann) magnetic field is generated around the laser axis by poloidal current loops induced by the non-uniform $\nabla p_{e}$ force. The third term on the right hand side is the Hall term that has pinching effects on the magnetic fields. To prevent the overestimation of self-generated magnetic fields at the edge of the coronal plasma, the magnetic-field source and Hall terms were calculated using limited scale lengths \cite{Igumenshchev2014}. This was implemented by replacing the grid size $\Delta x$ in the spatial derivatives of these terms by max($\Delta x$, $\epsilon l_{e}$), where $l_{e}$ is the electron mean free path and $\epsilon$ is a parameter on the order of unity. $\mathbf{R_{T}}$ and $\mathbf{R_{u}}$ were calculated with the full Braginskii transport coefficients, including the Nernst term and anisotropic magnetic resistivity \cite{Braginskii1965}. Flux-limited Spitzer-H\"{a}rm heat transport was used to calculate the electron heat flux \cite{Malone1975}, with cross-field, Righi-Leduc heat flow \cite{Kho1985} included in the electron energy equation. The data were modeled using the same target parameters and temporal history of the laser power that were used in the experiment. 

\begin{figure}[t]
\includegraphics{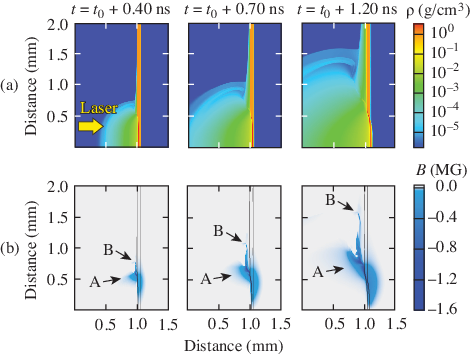}
\caption{\label{fig3}(color online). Simulated density profiles (a) and spontaneous magnetic field distributions (b) at times $t=t_{0}+0.40$ ns, $t=t_{0}+0.70$ ns, and $t=t_{0}+1.20$ ns. The calculations are axisymmetric about the horizontal axis. The negative sign indicates magnetic fields out of the page. Magnetic fields at the edge of the laser-focal region (A) and the edge of the coronal plasma (B) are shown (see Sec. 1 of Supplemental Material for details about regions A and B \cite{Supp1}).}
\end{figure}

Figure~\ref{fig3}(a) shows the calculated target-density profiles from DRACO at times $t=t_{0}+0.40$ ns, $t=t_{0}+0.70$ ns, and $t=t_{0}+1.20$ ns. The laser-ablated plasma accelerated the central part of the foil toward the right. The driven foil had a transverse size comparable with the laser focal spot and was not affected by laser burn through \cite{Li2007} or Rayleigh-Taylor instability growth over this period, becoming slightly bow-shaped in time \cite{Gao2012,Gao2013}. 

Figure~\ref{fig3}(b) shows the predicted magnetic field distributions for the same three interaction times. Overlaid on top of each image is the location of the driven target at that time. The calculations show the largest magnetic fields have MG-level magnitude and bound the laser-focal region (region A), expanding across the target surface at $\sim$0.3 $\times$ 10$^{8}$ cm$/$s. A second source of magnetic field is generated at the edge of the coronal plasma, close to the target surface where non-collinear density and temperature gradients also exist (region B). These magnetic fields, smaller in magnitude than those at the edge of the laser-focal region, expand with the coronal plasma at the plasma sound speed $\sim$1 $\times$ 10$^{8}$ cm$/$s. In between regions A and B, the magnetic fields expand radially along the target surface.




Synthetic proton radiographs were generated for direct comparison of the model predictions with the measured proton radiographs [Fig.~\ref{fig2}(b)]. The electromagnetic field distributions from the DRACO calculations were post processed with a proton tracking code that used the same proton radiography geometry as the experiment. In these calculations, the force deflecting the protons was $e(\mathbf{v} \times {\bf{B}})/c - (\nabla p_{e}-\mathbf{R_{T}})/n_{e}$. A simulated detector plane monitored the spatial distribution of the accumulated protons as a function of time, generating a time series of proton fluence images. In generating the synthetic proton radiographs, the film response was taken into account \cite{Hey2008}. The effect of collisional scattering and stopping in the experiment was small and not included in the proton tracking calculations \cite{Gao2012}.

In general, the model calculations reproduce the main features that were observed in the data. At the coronal plasma front (region II, Fig.~\ref{fig2}), electric fields deflected incident protons while magnetic fields focused them. These two effects caused a local proton-flux deficit and a light-colored ring on the detector where the protons would have been located if they were not deflected. This light-colored ring shows the location of the outermost coronal plasma, which expands at about the local plasma sound speed. Getting the magnitude of the fields correct in this region requires the consideration of kinetic effects that are not included in the DRACO simulations. Given the approximations used, region II in Fig.~\ref{fig2} can still be identified in the synthetic proton radiographs and used to track the evolution of the coronal plasma.

\begin{figure}[t]
\includegraphics{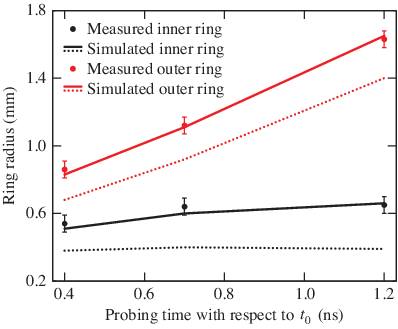}
\caption{\label{fig5} (color online). Comparison of the experimentally measured inner and outer ring radii with DRACO calculations post processed with a proton tracking code. Calculated inner (black) and outer (red) ring radii with (solid line) and without (dotted line) Nernst advection and Righi-Leduc heat flow included in the DRACO model.}
\end{figure}

Magnetic fields generated in the laser-focal region [region A, Fig.~\ref{fig3}(b)] focused incident protons, creating the inner dark ring in Fig.~\ref{fig2} (region I) by locally increasing the proton flux at the detector. When the electric fields were turned off in the simulations, regions I and II in Fig.~\ref{fig2} were still observed in the synthetic radiographs, confirming the importance of magnetic field effects in these regions (see Sec. 1 of Supplemental Material \cite{Supp1}). 

In time, multiple ring structures appeared in the synthetic radiographs. The ring structures at intermediate radii were not observed in the experiment. Small-scale electromagnetic fields in the corona likely masked these features in the experiment. At time $t=t_{0}+0.70$ ns, the data show flower-like patterns in the corona in between regions I and II. The scale length of these modulations grew and filamentary structures appeared at later times. Various instabilities such as the magnetothermal instability \cite{Haines1986} could develop in the corona and generate these electromagnetic fields, contributing to a less ordered deflection of the proton beam in this region. The MHD model does not describe these processes.



Figure~\ref{fig5} compares the evolution of the measured and predicted inner and outer rings. The measured ring radii were determined from an angular average. The outer ring radius at a given time was independent of the proton energy. The inner ring radius at a given time varied as a function of proton energy because different proton energies had different deflection angles. Ring radii based on post processed DRACO calculations are shown with and without the Nernst term and Righi-Leduc heat flow included in the model (see Sec. 2 of Supplemental Material \cite{Supp1}). 

When the Nernst term and Righi-Leduc heat flow were included in the DRACO calculations, larger inner and outer ring radii were predicted. The inner ring radius is larger because the Nernst term more effectively convects magnetic fields outward along the target surface. The outer ring radius is larger because redirected heat flow increases the coronal-plasma temperature near the target surface in the lateral direction, causing the plasma to expand faster. To reproduce the data and correctly account for the measured magnetic field generation and transport, all the terms in Eq.~(\ref{eq:draco}) had to be included in the DRACO model (other than the Hall term which has a small effect under these conditions). 

In summary, magnetic field generation and transport were studied at the surface of a laser-irradiated planar solid target using ultrafast proton radiography. The data show magnetic fields concentrated at the edge of the laser-focal region, well within the expanding coronal plasma. The measurements are in good agreement with a time series of synthetic proton radiographs post processed using field profiles calculated using the 2D resistive MHD code DRACO. The Biermann battery source, fluid and Nernst advection, resistive magnetic diffusion, and Righi-Leduc heat flow had to be taken into account in the calculations to reproduce the experimental results. This work provides significant insight into the generation and transport of Biermann fields in laser-produced plasmas, particularly those used in laser-driven magnetic reconnection and laboratory astrophysics experiments.

This work was supported by the U.S. Department of Energy Office of Inertial Confinement Fusion under Cooperative Agreement No. DE-FC52-08NA28302, the University of Rochester, and the New York State Energy Research and Development Authority. The support of DOE does not constitute an endorsement by DOE of the views expressed in this article. 


\bibliography{1MAIN}

\end{document}